\documentclass[prb,aps,twocolumn,floatfix,longbibliography]{revtex4-2}
\usepackage{hyperref}
\usepackage[utf8]{inputenc}
\usepackage{natbib}
\usepackage{graphicx}
\usepackage{color}
\usepackage[tbtags]{amsmath}
\usepackage{amssymb}
\usepackage{gensymb}
\usepackage{float}

\begin{document}

\title{Spectroscopic signatures of excitonic order effect on quantum spin Hall edge states}

\author{Micha{\l} Papaj}
\affiliation{Department of Physics, University of California, Berkeley, CA 94720, USA}

\begin{abstract}
One of the proposed ground states of monolayer WTe$_2$, a two-dimensional topological insulator, is an excitonic condensate. However, time-reversal preserving and breaking solutions are competing at the mean-field level of analysis, and it is unclear which condensate, if any, is realized in nature. In this work we analyze the experimental signatures that allow to provide evidence for the excitonic ground states and to distinguish between the two types of condensates using scanning tunneling microscopy. We provide clear experimental signatures in local mapping and quasiparticle interference patterns that visualize the expected changes in local charge and spin density, and characterize allowed backscattering processes in the presence of impurities and disorder. Our results will thus help in the determination of the nature of the WTe$_2$ ground state.
\end{abstract}

\maketitle

\section{Introduction}

One of the most important characteristics of the quantum spin Hall effect state \cite{kaneTopologicalOrderQuantum2005a, kaneQuantumSpinHall2005b, bernevigQuantumSpinHall2006a, bernevigQuantumSpinHall2006b, qiTopologicalInsulatorsSuperconductors2011, hasanColloquiumTopologicalInsulators2010}, or two-dimensional topological insulator (2D TI), is the presence of helical edge states at the boundary with a region of an opposite topological index. These edge states are in theory responsible for the quantized sample conductance of 2 e$^2$/h as long as the chemical potential is tuned into the bulk band gap of the material and the time-reversal symmetry is preserved in the system. However, in the first experimentally discovered 2D TI, the HgTe/CdTe quantum wells \cite{bernevigQuantumSpinHall2006a, konigQuantumSpinHall2007a, rothNonlocalTransportQuantum2009}, the quality of quantization was far from satisfactory. While the situation has since improved, many theoretical proposals were given in order to explain this breakdown of conductance quantization \cite{vayrynenHelicalEdgeResistance2013, vayrynenResistanceHelicalEdges2014, vayrynenNoiseInducedBackscatteringQuantum2018, hsuEffectsNuclearSpins2018, hsuNuclearspininducedLocalizationEdge2017, wuHelicalLiquidEdge2006, xuStabilityQuantumSpin2006, stromEdgeDynamicsQuantum2010, geisslerRandomRashbaSpinorbit2014, kainarisConductivityGenericHelical2014, xieTopologicalProtectionRandom2016, kharitonovBackscatteringHelicalLiquid2017, delmaestroBackscatteringHelicalEdge2013, crepinRenormalizationGroupApproach2012, budichPhononInducedBackscatteringHelical2012, groenendijkFundamentalLimitsHelical2018}. This deficiency in early 2D TIs prompted the search for new material platforms \cite{knezEvidenceHelicalEdge2011, knezObservationEdgeTransport2014, duRobustHelicalEdge2015, liObservationHelicalLuttinger2015, yangSpatialEnergyDistribution2012, culcerTransportTwodimensionalTopological2020} and one of the recently discovered promising systems for quantum spin Hall effect are monolayers of WTe$_2$ \cite{qianQuantumSpinHall2014, feiEdgeConductionMonolayer2017, tangQuantumSpinHall2017a, wuObservationQuantumSpin2018a, liQuantumSpinHall2020, jiaDirectVisualizationTwodimensional2017, pengObservationTopologicalStates2017}. Apart from the topological properties, this material also offers an exciting playground for exploring correlated effects as it exhibits superconductivity when doped \cite{fatemiElectricallyTunableLowdensity2018}. Unfortunately, even in this system the breakdown of quantization for edge lengths longer than 100 nm was observed \cite{wuObservationQuantumSpin2018a}. Therefore, it is interesting to determine the possible cause for this behavior so that the future samples could be optimized to minimize the effect of the leading mechanism.

Another interesting feature of WTe$_2$ is the proposed existence of excitonic condensate in the ground state of this system \cite{varsanoMonolayerTransitionmetalDichalcogenide2020, sunEvidenceEquilibriumExciton2022, jiaEvidenceMonolayerExcitonic2022}, which enables studying the interplay of excitonic condensation and topological properties \cite{pikulinInterplayExcitonCondensation2014, duEvidenceTopologicalExcitonic2017, blasonExcitonTopologyCondensation2020, paulInterplayQuantumSpin2022, paulInterplayQuantumSpin2022a, wangExcitonicTopologicalOrder2023}. Such proposals are based on the possible coexistence of electron and hole pockets in the non-interacting band structure as predicted by some of the first principles band structure calculations. The hole pocket is centered around the $\Gamma$ point, while the two electron pockets are placed symmetrically at $\pm q_c$ along the high symmetry direction $\Gamma - X$ in the material's Brillouin zone. The screened Coulomb interaction then leads to the formation of excitons with finite momentum $q_c$ as evidenced by the mean-field calculations based on the Hartree-Fock approach. However, depending on the way the self-consistent calculation is performed, one may arrive either at time-reversal preserving or breaking solutions \cite{kwanTheoryCompetingExcitonic2021, wangBreakdownHelicalEdge2023}. Even though the energy of the ground state that spontaneously breaks time-reversal is lower, as the energy difference is small it is difficult to unequivocally decide which state will be realized in reality. It is therefore an interesting problem to define the signatures of each of the states and to provide experimental procedure that could distinguish between the two. One possible approach stems from the impact of the excitonic condensate on the transport properties of the system. In particular, it can be shown that time-reversal breaking ground state has the form of a spin spiral, which enables backscattering within the same edge. This in turn leads to the breakdown of conductance quantization when the scattering off the spin spiral and non-magnetic disorder is combined. Quantum transport simulations reveal that such a combination can be a possible explanation of the observed transport data in WTe$_2$ \cite{wangBreakdownHelicalEdge2023}. If more careful measurements of temperature and edge length scaling are performed, the data could also be compared to the predictions based on Luttinger liquid model of the edge states when affected by the spin spiral \cite{hsuNuclearspininducedLocalizationEdge2017, hsuEffectsNuclearSpins2018, wangBreakdownHelicalEdge2023}. However, it is also interesting to study how the excitonic condensate can be visualized using local probes such as scanning tunneling microscope (STM).

In this work, we analyze the local signatures of the excitonic condensate in WTe$_2$ using the topological insulator edge states as a probe. We consider both the time-reversal preserving and breaking states, which correspond to charge-density-wave-like and spin spiral states, respectively. Local density of states and local spin density already reveal significant difference between the two classes of states, as only the time-reversal preserving state displays spatial modulation of the edge state, while the edge state in the presence of the spin spiral excitonic condensate remains uniform. On the other hand, there is no net spin density in the charge density wave case, but a clear modulation of spin density around the edge exists in the spin spiral case. However, even more significant difference is observed when quasiparticle interference (QPI) pattern is studied. In the presence of impurities and disorder, when the ground state is the spin spiral, the backscattering within the same edge is allowed and thus a clear dispersive signal is observed within the bulk gap when it is scanned over. In contrast, the charge density wave state does not show any signs of backscattering and QPI instead helps to visualize the edge state modulation of constant spatial period throughout the bulk gap. With both local mapping, quasiparticle interference patterns, and transport properties combined, our discussion enables visualization and differentiation between various possible excitonic states, which should help to experimentally clarify the nature of the ground state in WTe$_2$.

\section{Models}

\begin{figure}
    \centering
    \includegraphics[width=0.99\columnwidth]{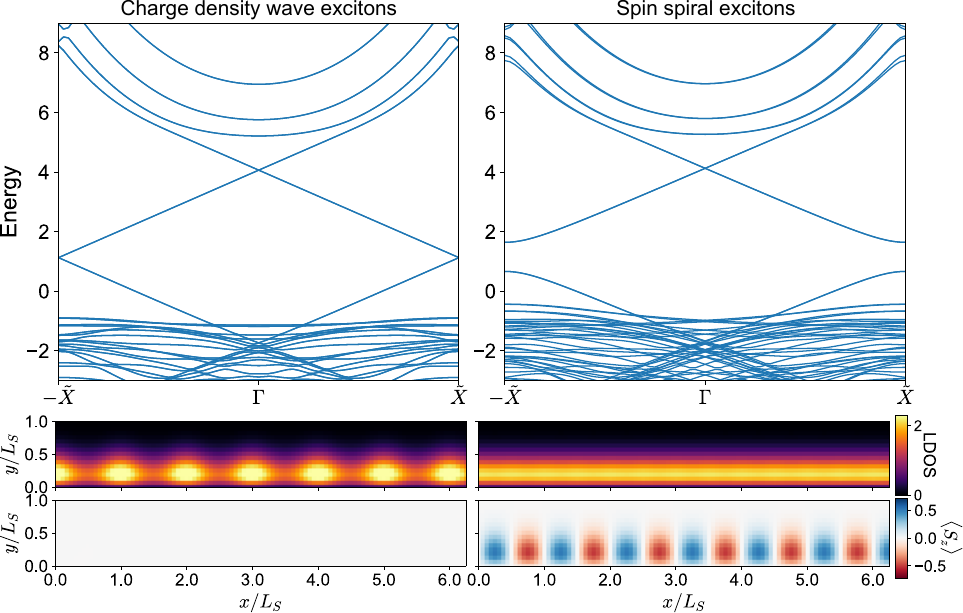}
    \caption{Comparison of two types of oscillating states within BHZ-based toy model. Upper panels show transverse subbands of the quasi-1D ribbon, while lower panels show local density of states and local spin density in the vicinity of the ribbon edge, calculated at energy $E=3$ in all cases.}
    \label{fig:ldos_BHZ}
\end{figure}

To investigate the impact of the excitonic states on the the edge states of quantum spin Hall effect we will use two different models. The first one is the very simple application of Bernevig-Hughes-Zhang (BHZ) model \cite{bernevigQuantumSpinHall2006a}, first used to describe HgTe/CdTe quantum wells, with effective terms that describe either charge density or spin spiral modulation observed in Hartree-Fock calculations for WTe$_2$. While this model may be overly simplistic and it will not be used with realistic material parameters, it can nevertheless be useful to illustrate some of the general impact of the spin spiral on the edge states without introducing additional complications that could obscure the interpretation. This simplification, which also avoids using Hartree-Fock to obtain the excitonic order parameter self-consistently allows to use larger system sizes with a greater spatial resolution. The Hamiltonian for this model is therefore given by:
\begin{align}
   H_{\mathrm{BHZ}}(\mathbf{k}) &= A (k_x s_z \sigma_x + k_y s_0 \sigma_y) + B\, k^2 s_0 \sigma_z + D\, k^2 s_0 \sigma_0 \notag \\ &+ M s_0 \sigma_z  + S \left[\cos(q_s x)s_x \sigma_0 + \sin(q_s x)s_z \sigma_0\right] \notag \\ &+ V \cos (q_s x) s_0 \sigma_0
   \label{eq:ham_BHZ}
\end{align}
where $k^2 = k_x^2 +k_y^2$,  $s_i$ and $\sigma_i$ are Pauli matrices in spin and orbital spaces, $A$, $B$, $D$, and $M$ are BHZ model parameters, $S$ determines the strength of the spin spiral effective magnetic field, while $V$ gives the magnitude of charge density wave-like excitonic condensate for comparison purposes. The period of the spatial modulation $L_S = 2\pi/q_s$ is chosen such that the gap that opens up due to coupling of states with momenta separated by $q_s$ still lies within the bulk band gap. In all of the calculations the parameters used are $A=2.5$, $B=-1.0$, $D=-0.8$, $M=-5$, and $S=1$ or $V=1$ for spin spiral and charge density wave states, respectively. The parameters for the BHZ model were chosen so that the helical edge state has a simple exponential decay into the bulk, instead of additional oscillating pattern \cite{papajConductanceOscillationsQuantum2016}. When discretizing this model on a lattice we use lattice constant $a_L = 0.1$ and the modulation period is then $L_S = 16\,a_L$.

To make more realistic material predictions we employ the second model, which is based on the effective low energy description of WTe$_2$. Its parameters are chosen to reproduce the first-principles band structure in the semimetallic phase of this material. The model consists of two orbitals and two spin orientations with the Hamiltonian \cite{jiaEvidenceMonolayerExcitonic2022}:
\begin{align}
H_{\mathrm{WTe}_2}(\mathbf{k}) = &\left(a k_x^2 + b k_x^4 + 2 b k_x^2 k_y^2 + b_y k_y^4 + \frac{\delta}{2}\right) I_d \notag \\ &+ \left(-\frac{k^2}{2m} - \frac{\delta}{2}\right) I_p + v_x k_x \tau_x s_y + v_y k_y \tau_y s_0 
\label{eq:hamiltonian}
\end{align}
where $\tau_i$ and $s_i$ are Pauli matrices in $p$, $d$ orbitals and spin spaces, respectively, $I_d = (\tau_0 + \tau_z)/2\, s_0$ and $I_p = (\tau_0 - \tau_z)/2\, s_0$ are the identity matrices corresponding to the $d$ and $p$ orbitals, and $v_x$ and $v_y$ characterize the spin-orbit coupling. The parameter values we use in the calculations are as determined in Ref.~\cite{jiaEvidenceMonolayerExcitonic2022} are $a=-3, b=18, b_y=40, \delta=-0.9, m=0.03, v_x=0.5, v_y=3$, where all the energies are expressed in eV and lengths in {\AA}. The one exception is the parameter $b_y$, which was changed to have correct low energy behavior within the whole Brillouin zone when the continuum model is discretized on a lattice. With such a model, we performed Hartree-Fock mean field calculations to obtain the excitonic order parameters that are also later represented in the discretized lattice version. Further details of the self-consistent calculation and the transport characterization of the resulting state with excitonic condensation were presented in Ref.~\cite{wangBreakdownHelicalEdge2023}.

In the case of both models to perform the calculations we use the lattice representation of the continuum models obtained through the finite-difference method. This will allows us to easily introduce local changes to the Hamiltonian, either by placing a single impurity at the edge or by including random disorder in the whole sample. All of the lattice simulations were performed using Kwant package \cite{grothKwantSoftwarePackage2014b}. The discretized models were studied in a quasi-one-dimensional ribbons of finite width and with hard walls boundary condition. The ribbons were oriented along the $x$ direction, which is the direction of the exciton-induced modulation. This means that momentum along $x$ direction remains a good quantum number after folding of the bands (associating states with momenta $k_x$ and $k_x+q_c$) and allows for characterization of the states within the ribbon by the transverse subbands with their own band structure.

Whenever we discuss placing an impurity at position $\mathbf{r}_0$, we include it as on-site potential given by:
\begin{equation}
    V_\mathrm{imp}(\mathbf{r}) = V_0 s_0 \sigma_0 \delta_{\mathbf{r},\mathbf{r}_0}
\end{equation}
where $V_0$ is the strength of scalar impurity. When including random disorder in the sample, we are choosing values $U_\mathrm{dis}(\mathbf{r})$ from the uniform random distribution in the range $[-U_0/2, U_0/2]$, where $U_0$ is the disorder strength. The disorder enters the Hamiltonian as on-site potential as well, similarly to the isolated impurities.

\section{Local density of states in the presence of excitonic states}

\begin{figure}
    \centering
    \includegraphics[width=0.99\columnwidth]{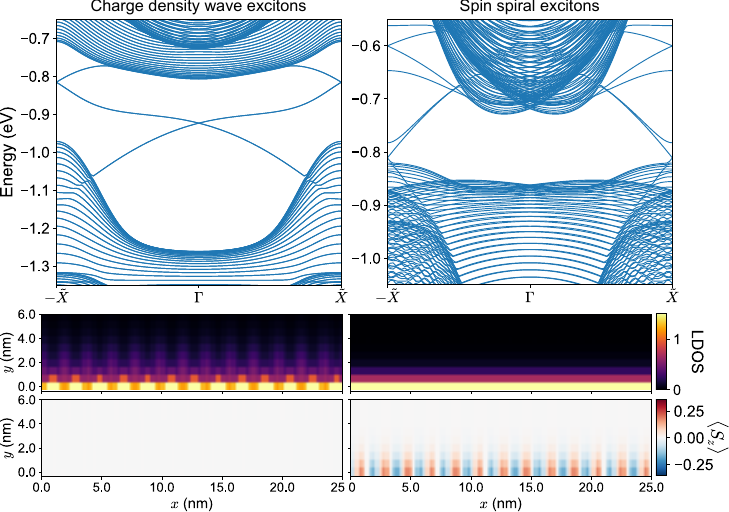}
    \caption{Comparison of two types of excitonic states within WTe$_2$ model. Upper panels show transverse subbands of the quasi-1D ribbon, while lower panels show local density of states and local spin density in the vicinity of the ribbon edge, calculated at $E=-0.85\,\mathrm{eV}$ and $E=-0.75\,\mathrm{eV}$ in the charge density wave and spin spiral cases, respectively.}
    \label{fig:ldos_WTe2}
\end{figure}

We begin with analyzing the effective BHZ model with either spin spiral or charge density wave introduced via non-zero $S$ and $V$ parameters in Eq.~\eqref{eq:ham_BHZ}. The spectrum of 1D transverse subbands for this model is presented in Fig.~\ref{fig:ldos_BHZ} in two cases: the charge density wave excitons ($V=1$, $S=0$) and spin spiral excitons ($V=0$, $S=1$). The energy windows of both figure panels are focused around the bulk band gap of BHZ model, with both conduction and valence bands visible. As we are performing our calculations in the quasi-1D ribbon geometry and the choice of parameters places the BHZ model in the topological phase, in both cases helical edge states are prominently visible throughout the gap. However, since the charge density wave excitons do not break the time-reversal symmetry the edge states are gapless, even though the density wave couples states separated by $q_s$ and causes band folding. In contrast, the spin spiral state results in mini-gap opening at the points of band folding, since the spin spiral enables hybridization of spin-up and spin-down states. Notably, even though the spin spiral breaks time reversal symmetry, the gap doesn't open at the Dirac point at $k_x = 0$, since when integrated over the whole period of the spiral, there is no net magnetic moment and time-reversal symmetry is preserved on average. In general however, outside of the vicinity of the zone boundary, in both cases the spectra remain similar.

We can now investigate how both of the excitonic states affect the local density of states (LDOS) for energies within the bulk gap as presented at the bottom of Fig.~\ref{fig:ldos_BHZ}. As expected, the states are localized at the hard wall boundaries of the ribbon. However, the appearance of the edge states is radically different in both analyzed cases. The charge density wave excitons impose their modulation on helical edge states, which results in fringing pattern observed in LDOS. In contrast, the spin spiral leaves the density of states at the edges uniform as it doesn't have any charge density modulation itself. Nevertheless, the oscillating pattern is revealed when net spin moment density $\langle S_z \rangle$ is inspected. Even though the total density of the edge states is uniform, the time-reversal breaking character of the spin spiral results in periodic oscillation of spin density with the wave vector $q_s$. On the other hand, such oscillation is absent in time-reversal preserving charge density wave excitonic condensate. Therefore, already at the level of local density of states the two types of excitonic condensate can be distinguished, which could be achieved using STM, in particular with spin-polarized tips.

\begin{figure*}
    \centering
    \includegraphics[width=0.999\linewidth]{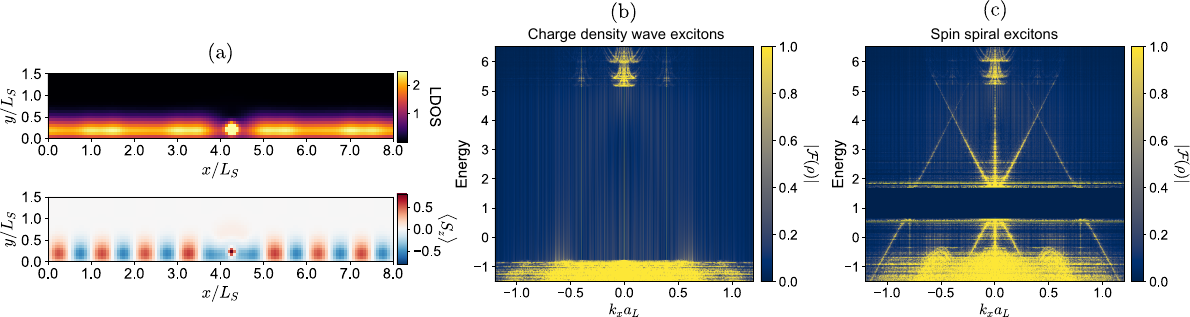}
    \caption{The effects of impurity and disorder scattering on the edge states in the presence of excitonic condensates in the BHZ model. (a) Local density of states and local spin density in the spin spiral case in the vicinity of an impurity with $V_0=40$. Backscattering enabled by time-reversal symmetry breaking introduces modulation into the edge state. (b) QPI spectrum due to random on-site disorder with $U_0=5$ for the charge density wave excitons. Within the bulk gap signal consists of density of states of gapless helical edge states and the periodic modulation of the excitonic charge density (very sharp and bright lines at $k_x a_L \approx \pm 0.4$). (c) QPI spectrum due to random on-site disorder with $U_0=5$ for the spin spiral excitons. Within the bulk gap we observe dispersing signal resulting from backscattering between the states at the same edge and mini-gap that opens for states with momenta given by the spatial frequency of spin spiral.}
    \label{fig:qpi_BHZ}
\end{figure*}

We can now perform a similar analysis for the true excitonic states obtained within the Hartree-Fock mean field approximation for the model of WTe$_2$. We are again considering two scenarios: time-reversal preserving and time-reversal breaking. Which of these is realized depends on the constraints put on the self-consistent calculation: if we start the mean field calculation from an initial state that preserves time-reversal symmetry and we make sure that TR-breaking perturbation does not get generated from numerical effects such as round-off errors, the calculation will converge to an excitonic condensate which preserves time reversal symmetry and leads to a charge density wave-like state. The subband spectrum of a ribbon with such a state is presented in Fig.~\ref{fig:ldos_WTe2}. As the TR symmetry is present, even though the bulk gap is due to the formation of an excitonic condensate, the helical edge states of the two-dimensional topological insulator phase are prominently displayed. However, as the excitonic condensate has finite center-of-mass momentum $q_c$ which is defined by the momentum space separation of the electron and hole pockets of WTe$_2$, the charge density wave state leads to the modulation of the local density of states at the edge, similarly to the previously considered toy model. This modulation is clearly visible in the LDOS plot shown below the spectrum, even though due to the limitation of the Hartree-Fock calculation we are limited to the density wave period of 7 lattice sites. At the same time, there is no spin density modulation visible in this case as the excitonic states are spin-degenerate.

If on the other hand we allow for TR symmetry breaking in the self-consistent calculation, we arrive at the spin spiral state, which in general has lower energy than the TR-preserving ground state. The data for the spin spiral state is presented in the right column of Fig.~\ref{fig:ldos_WTe2}. The subband spectrum still displays a bulk gap, but the spin degeneracy is lifted and the state can no longer be classified as a true topological insulator due to TR-breaking. Still, the remnants of the helical edge states are visible within the gap due to the relative weakness of the TR-breaking perturbation introduced by the excitonic spin spiral. The local density of states and the local spin density again display features that were prominent in the toy model, albeit with a lower spatial resolution. The LDOS plot shows that the edge states in this case are uniform, with no sign of modulation caused by the finite momentum excitonic condensate. Due to TR-breaking, this modulation is visible in the spin density, however. The period of this modulation, similarly to TR-preserving case is only dependent on the momentum separation of electron and hole pockets that undergo condensation and is thus independent of energy at which the measurement is performed. This analysis demonstrates that the TR-preserving and breaking excitonic condensates can be visualized and distinguished already using the spatial pattern at the sample edges, preferably using spin-polarized STM technique. This is true both in the toy model, as well as in more realistic WTe$_2$ model that employs Hartree-Fock order parameters to describe the excitonic condensate.

\section{Quasiparticle interference patterns}

Apart from the gap opening and the spin density modulation, breaking of time-reversal symmetry allows for an additional significant effect: breakdown of topological protection of the helical edge states, leading to the possibility of backscattering within the same edge. To study this phenomenon we can employ the technique of quasiparticle interference, in which the Fourier transform of the local density of states can reveal the allowed scattering processes between various states at the Fermi energy. We first visualize the modulation introduced by scattering by plotting the local density of states in a system in which an impurity was placed in the vicinity of the edge state. This is shown for the BHZ model in Fig.~\ref{fig:qpi_BHZ}(a) in the spin spiral case. While in absence of defects the density of states is uniform along the edge, the inclusion of an impurity introduces long range modulation into the edge state, signifying that backscattering between spin-up and spin-down states occurs. In contrast, no major long range perturbation occurs in time-reversal preserving situation. The same modulation is also visible in the local spin density, although less clearly than in the total density of states. We can now characterize the period of the backscattering-related modulation using the Fourier transform.

In Fig.~\ref{fig:qpi_BHZ} we present scans of Fourier transforms of the local density of states $\rho(x)$ along the edge of the ribbon as a function of momentum $k_x$ and the energy. In the time-reversal preserving case, Fig.~\ref{fig:qpi_BHZ}(b), at low and high energies we see signal coming from valence and conduction subbands, with stronger signal coming from the valence band due to the larger number of subbands and thus larger density of states available for scattering. Inside of the bulk gap, however, we don't observe any scattering signal dispersing with energy, demonstrating that the topological edge states are protected from backscattering since time-reversal symmetry is preserved. This remains true even though the excitonic condensate is present in the system. What we observe is zero momentum signal corresponding to the local density of states at a given energy, which indicates the presence of the edge state. We also observe two symmetrically placed lines at finite momenta ($k_x a_L = \approx \pm 0.4$) corresponding to the spatial frequency of charge density wave modulation coming from the excitonic condensate. These lines are very sharp and bright, further signifying their robustness against non-magnetic disorder. Since the period of the modulation that results from the excitonic order parameter is independent of energy, the lines are not dispersive and are purely vertical in the quasiparticle interference spectrum.

However, the picture observed in the spin spiral case, presented in Fig.~\ref{fig:qpi_BHZ}(c), is radically different from the time-reversal preserving case. Since the local density of states in the unperturbed time-reversal breaking scenario is uniform, there are no vertical lines visible inside of the gap at momenta corresponding to the spin spiral spatial frequency. In their place instead we can clearly see dispersive signal that results from backscattering between the two counterpropagating edge states at the given edge. This behavior of the QPI signal mimicks the energy dispersion of the edge state shown previously in the subband spectrum. We also observe the mini-gap that opens up in the spectrum due to the coupling of states separated in momentum space by the spatial periodicity of the spin spiral. These combined features provide a clear signature of the presence of the spin spiral excitonic condensate when diagnosed via the behavior of the edge states.

\begin{figure}[!t]
    \centering
    \includegraphics[width=0.99\columnwidth]{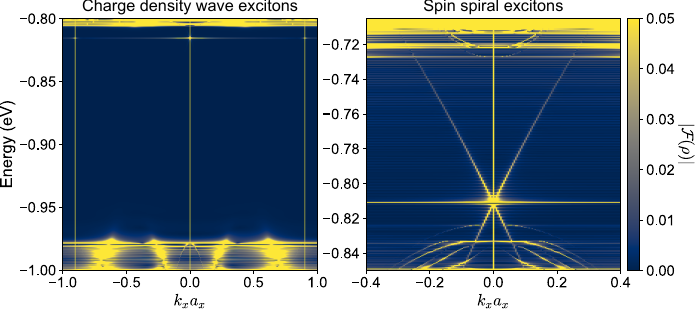}
    \caption{The effects of single impurity scattering with $V_0=1\,\mathrm{eV}$ on the edge states in the presence of excitonic condensates in WTe$_2$ model. The time-reversal preserving state (left panel) shows charge modulation of constant period throughout the bulk gap, while the time-reversal breaking state (right panel) displays dispersing signal resulting from backscattering within the same edge.}
    \label{fig:qpi_WTe2}
\end{figure}

To verify the predictions about the structure of quasiparticle interference patterns obtained from the toy model, we again perform a similar analysis for the more realistic WTe$_2$ model with Hartree-Fock mean-field order parameters. However, due to the smaller spatial resolution of this model and the short decay length of the edge states (a few lattice sites into the bulk of the ribbon), we limit our consideration to patterns resulting from a single impurity placed in the vicinity of the edge. Such QPI spectra are presented in Fig.~\ref{fig:qpi_WTe2} for both types of excitonic order that can result from the self-consistent calculation. In the case of charge density wave excitons obtained while enforcing time-reversal symmetry in the ground state, the QPI also displays consistent presence of the helical edge states as the signal at $k_x=0$ throughout the bulk band gap. We also note clear signal coming from the spatial oscillation of the edge density of states with periodicty given by $q_c$. Other than that, no other signal is present within the gap, signifying that even the proper excitonic order parameter, with multiple components that couple all the spin and orbital degrees of freedom does not introduce any form of backscattering within the same edge of the ribbon.

Similarly to the toy model, the QPI pattern is entirely different when the ground state of the system is the time-reversal breaking spin spiral. In that case the periodic modulation of the local density of states at the edge vanishes, leaving the oscillation visible only in the local spin density results. On the other hand, a new signal that is dispersive appears within the gap. As in the toy model case, the signal corresponds to the backscattering that is possible with the same edge after breaking time-reversal symmetry. We have verified this by checking that this signal does not change when the width of the ribbon (and thus the distance between the opposite edge states) is increased. The in-gap signal lines meet at $k_x=0$ for the energy for which the edge state dispersion in the subband spectrum is at the folding point corresponding to the half of the exciton condensate momentum $q_c$. The following linear dispersion of the signal lines reflects the linear dispersion of the remaining in-gap edge states also visible in the spectrum. Presence of this signal thus demonstrates that even when using a more realistic Hartree-Fock model with multiple different order parameters, we can still clearly distinguish the time-reversal preserving and breaking cases when their quasiparticle interference patterns are compared.

\section{Summary and outlook}

In summary, in this work we analyzed the STM spectroscopic signatures of the excitonic condensate in WTe$_2$. We demonstrated that the nature of the excitonic ground states can be distinguished between the two possible candidate states both by analyzing the real space pattern of local density of states and local spin density in the vicinity of the sample edges as well as by considering the quasiparticle interference patterns. One additional tuning parameter that could possibly be applied to further verify the existence of the excitonic condensate is the application of external magnetic field. If the ground state has spin spiral nature this will lead to the appearance of the previously absent charge density wave even in the clean edge. With these clear signatures, combined with the previously provided transport characteristics of the 2D topological insulator with the excitonic condensate ground state, it should be possible to experimentally verify whether WTe$_2$ indeed does realize such an excitonic ground state.

\begin{acknowledgements}
 We thank Tiancong Zhu and Michael Crommie for for helpful discussions. This material is based upon work supported by the U.S. Department of Energy, Office of Science, National Quantum Information Science Research Centers, Quantum Science Center.  M.P. received additional fellowship support from the Emergent Phenomena in Quantum Systems program of the Gordon and Betty Moore Foundation.
\end{acknowledgements}

\bibliography{STM_signature_spin_spiral}

\end{document}